\documentclass[aps,prl,twocolumn,groupedaddress,superscriptaddress,floatfix]{revtex4}

\usepackage{amsmath, amsthm, amssymb}
\usepackage{subfigure}
\usepackage{graphicx}
\usepackage{dsfont}

\usepackage{xcolor}
\definecolor{mydarkgreen}{rgb}{0.0,0.5,0.0}

\usepackage{afterpage}

\usepackage[linktocpage=true,
  colorlinks=true, 
  pdfborder={0 0 0},
  linkcolor=blue,
  citecolor=red,
  filecolor=yellow,
  urlcolor=mydarkgreen,
  bookmarks,
  pdftitle={RDMFT: Minimization},
  pdfauthor={Tim Baldsiefen},
 pdfkeywords={Reduced density matrix functional theory,Minimization},
]{hyperref} 

\newcommand{\mc}[1]{\mathcal{#1}}
\newcommand{\mf}[1]{\mathfrak{#1}}
\newcommand{\tb}[1]{\textbf{#1}}
\newcommand{\tr}[1]{\textup{tr}\{#1\}}
\newcommand{\eq}[1]{(\ref{#1})}

\newcommand{\ket}[1]{{|#1\rangle}}
\newcommand{\bra}[1]{{\langle#1|}}

\newcommand{\nn}{\nonumber}
\newcommand{\veps}{\varepsilon}

\newcommand{\abref}[1]{#1}

\begin{document}


\title{Minimization procedure in reduced density matrix functional theory by means of an effective noninteracting system}


\author{Tim Baldsiefen}
\affiliation{Institut f\"ur Theoretische Physik, Freie Universit\"at Berlin, Arnimallee 14, D-14195 Berlin, Germany}
\affiliation{Max-Planck-Institut f\"ur Mikrostrukturphysik, Weinberg 2, D-06112 Halle, Germany}
\author{E. K. U. Gross}
\affiliation{Max-Planck-Institut f\"ur Mikrostrukturphysik, Weinberg 2, D-06112 Halle, Germany}


\date{\today}

\begin{abstract}
In this work, we propose a self-consistent minimization procedure for functionals in reduced density matrix functional theory.  We introduce an effective noninteracting system at finite temperature which is capable of reproducing the groundstate one-reduced density matrix of an interacting system at zero temperature. By introducing the concept of a temperature tensor the minimization with respect to the occupation numbers is shown to be greatly improved.
\end{abstract}


\maketitle


\section{Introduction}
Since 1964, after the pioneering work of Hohenberg and Kohn \cite{Hohenberg_Kohn.1964}, density functional theory (DFT) became the standard tool for the calculation of groundstate (gs) properties of quantum-mechanical systems. There are, however, some physical problems which are difficult to address in the framework of DFT. These include the description of strongly correlated systems, such as the dissociation of closed shell molecules into open shell fragments, and the fundamental gap in Mott insulators. Recently, a promising alternative to DFT was introduced which showed success in various fields, ranging from small molecules \cite{Mueller.1984,Baerends.2001,Buijse_Baerends.2002,Gritsenko_Pernal_Baerends.2005,Helbig_al.2007,Marques_Lathiotakis.2008,Piris_al.2010,Helbig_Lathiotakis_Gross.2009} to infinite solids \cite{Baldsiefen.2010,Lathiothakis_al.2009,Sharma_al.2008}, including the difficult cases mentioned above. This method features the one-reduced density matrix (1RDM) as central variable and is called reduced density matrix functional theory (RDMFT). In the theoretical framework of RDMFT, the functional form of the kinetic as well as of the exchange energy are known exactly in terms of the 1RDM and only the correlation part of the two-particle interaction energy has to be approximated. However, a minimization of functionals in RDMFT is complicated by the fact that at zero temperature there is no noninteracting system reproducing the 1RDM of the interacting system. This is in contrast to DFT where the Kohn-Sham system \cite{Kohn_Sham.1965} allows for an efficient self-consistent minimization. Therefore, in RDMFT one usually resorts to direct minimization routines.

In the present work, we show that one can indeed construct a noninteracting system which reproduces a given 1RDM to arbitrary accuracy, if one considers this system to be in grand canoncial equilibrium at finite temperature. We therefore effectively model a zero-temperature interacting system by a finite-temperature noninteracting one. This allows one to construct a self-consistent Kohn-Sham minimization scheme for functionals in RDMFT.

Capitalizing on the freedom of choice for the temperature of the Kohn-Sham system, we will furthermore introduce the concept of a temperature tensor. This concept will later on be shown to greatly improve the performance of our minimization procedure.

We will then argue, why the energy value in a numerical minimization of a RDMFT functional is not a good measure of convergence. As alternatives we introduce two convergence measures which rely solely on the functional derivative of the RDMFT functional w.r.t. the 1RDM.

Finally, we will investigate the performance of the new minimization scheme by applying a common RDMFT functional to LiH. It will be shown that the self-consistent scheme is very efficient and avoids conceptual difficulties prevalent in many other minimization procedures.

\section{Theoretical foundations}
In this work, we will consider systems governed by a Hamiltonian $\hat H$ consisting of the kinetic energy $\hat T$, the external one-particle potential $V$, and the two-particle interaction $\hat W$:
\begin{align}
  \hat H&=\hat T+\hat V+\hat W.
\end{align}
A quantum-mechanical system is generally described by a statistical density operator (SDO) $\hat D$ being a weighted sum of projection operators on the Hilbert space under consideration
\begin{align}\label{eq.sdo.0}
  \hat D&=\sum_iw_i\ket{\Psi_i}\bra{\Psi_i},\quad w_i\geq0,\quad\sum_iw_i=1.
\end{align}
The 1RDM $\gamma(x,x')$, corresponding to a particular SDO $\hat D$, is defined as
\begin{align}
  \gamma(x,x')&=\tr{\hat D\hat\psi^+(x')\hat\psi(x)}\label{def.1rdm},
\end{align}
where $\{\hat\Psi(x)\}$ are the common field operators and the variable $x$ denotes a combination of spacial coordinate $\tb r$ and spin index $\sigma$ ($x=(\tb r,\sigma)$). An integration over $x$ is therefore to be interpreted as an integration over $\tb r$ and a summation over $\sigma$. By construction, $\gamma(x,x')$ is hermitean and can therefore be written in spectral representation
\begin{align}
  \gamma(x,x')&=\sum_in_i\phi_i^*(x')\phi_i(x)\label{def.1rdm.2}.
\end{align}
The $\{\phi_i(x)\}$ are traditionally called the natural orbitals (NO) and the $\{n_i\}$ are the occupation numbers (ON) \cite{Loewdin.1955}. The conditions that ensure that a given $\gamma(x,x')$ is ensemble-N-representable, i.e. that it comes from a SDO of the form of Eq.\eq{eq.sdo.0}, are the following \cite{Coleman.1963}.
\begin{align}
  &0\leq n_i\leq1\label{eq.1rdm.en_rep.1}\\
  &\sum_in_i=N\label{eq.1rdm.en_rep.2}\\
  &\{\phi_i\} \textup{ is a complete orthonormal set}\label{eq.1rdm.en_rep.3}
\end{align}
The set of all ensemble-N-representable 1RDMs is given by
\begin{multline}
  \Gamma^N=\Big\{\gamma(x,x')\ \Big|\\
  \gamma(x,x') \textup{ fulfills Eqs.} \eq{eq.1rdm.en_rep.1},\eq{eq.1rdm.en_rep.2},\textup{ and}\eq{eq.1rdm.en_rep.3} \Big\}
\end{multline}
which is closed and convex.

Following from the theorems of Hohenberg and Kohn \cite{Hohenberg_Kohn.1964}, we know that one can formulate a functional theory of the 1RDM for the determination of the gs energy. It was Gilbert \cite{Gilbert.1975} who showed that this theoretical framework is also capable of describing systems subject to nonlocal external potentials, a task not possible via DFT. We have furthermore shown in \cite{Baldsiefen_al_1.2012} that this methodology can be extended to the case of quantum-mechanical systems in grand canonical equilibrium. 

The functional for the energy $E[\gamma]$ of the interacting and for the grand potential $\Omega_0[\gamma]$ of a noninteracting system in grand canonical equilibrium are given as
\begin{align}
  E[\gamma]&=T[\gamma]+V_{ext}[\gamma]+W[\gamma]\label{eq.e}\\
  \Omega_0[\gamma]&=T[\gamma]+V_{ext}[\gamma]-\mu N[\gamma]-1/\beta S_0[\gamma]\label{eq.gp-ni}
\end{align}
where
\begin{align}
  T[\gamma]&=\int dx'\lim_{x\rightarrow x'}\left(-\frac{\nabla^2}{2}\right)\gamma(x',x)\\
  V_{ext}[\gamma]&=\int dxdx'v_{ext}(x,x')\gamma(x',x)\\
  N[\gamma]&=\int dx\gamma(x,x)\\
  S_0[\gamma]&=-\sum_i(n_i\ln n_i+(1-n_i)\ln(1-n_i))\label{eq.ent.ni}.
\end{align}
The functional $W[\gamma]$ for the interaction contribution is not known exactly and has to be approximated in practice. The noninteracting grand potential can be written solely in terms of the one-particle eigenenergies and the ONs as
\begin{align}
  \Omega_0[\gamma]&=\sum_i\Big(n_i(\veps_i-\mu)+\nn\\
  &\hspace*{15mm}\frac{1}{\beta}(n_i\ln n_i+(1-n_i)\ln(1-n_i))\Big)\label{eq.num.gp-ni-exp}.
\end{align}

In the context of this work, the question of noninteracting (ni)-V-representability, i.e. the question which 1RDMs correspond to a groundstate or equilibrium of a noninteracting system, will become important. The sets of all zero-temperature ni-V-representable and finite-temperature ni-V-representable 1RDMs will be denoted by $\Gamma^V_0$ and $\Gamma^V_T$, respectively. In the case of zero temperature a nondegenerate system assumes a pure groundstate and the corresponding noninteracting 1RDM will be idempotent. Therefore, $\Gamma^V_0$ is on the boundary of $\Gamma^N$. We have shown in \cite{Baldsiefen_al_1.2012} that the gs-1RDM of a Coulomb system is in the interior of $\Gamma^N$ and, therefore, we cannot use a noninteracting system at zero temperature to find the minimum of an RDMFT functional. In simple terms: The 1RDM of interacting particles is never idempotent and, hence, it cannot be represented as the 1RDM of a Kohn-Sham-type noninteracting system at zero temperature. At finite temperature, however, for a noninteracting system with one-particle eigenvalues $\{\varepsilon_i\}$, the ONs are given by the Fermi-Dirac distribution \cite{Dirac.1926} which can easily be inverted:
\begin{align}
  n_i&=\frac{1}{e^{\beta(\varepsilon_i-\mu)}+1}\\
  \varepsilon_i-\mu&=\frac{1}{\beta}\ln\left(\frac{1-n_i}{n_i}\right)\label{eq.ei.ni}.
\end{align}
This implies that all 1RDMs in the interior of $\Gamma^N$ are in $\Gamma^V_T$. Therefore, for every 1RDM in $\Gamma^N$ there is a 1RDM from $\Gamma^V_T$ arbitrarily close to it which allows the utilization of a noninteracting system in grand canonical equilibrium in a self-consistent minimization scheme. We emphasize the term ``arbitrarily close'' because pinned ONs (i.e. 0 or 1) cannot be reproduced by a system at finite temperature (see Eq. \eq{eq.ei.ni}), but every ON arbitrarily close to 0 or 1 can. The error introduced by these pinned states therefore becomes arbitrarily small.

\section{Self-consistent minimization}
The biggest stumbling stone in the numerical minimization of RDMFT functionals is the incorporation of the auxiliary constraints on the ONs and NOs of the 1RDM. These are particle number conservation $\sum n_i=N$, the fermionic constraint $0\leq n_i\leq1$, and most importantly, the orthonormality constraints of the NOs. Usually, the orthonormality of the \abref{NO}s will be enforced by applying an orthonomalization algorithm to the \abref{NO}s after they have been modified, using the information provided by the functional derivatives $\delta E[\gamma]/\delta\phi_i$. These orthonormalization procedures can change several orbitals quite significantly which can lead to a slow convergence of the minimization routines.

The main idea of a self-consistent minimization scheme is now to approximate the energy surface $E[\gamma]$ by a simpler one whose minimum, incorporating all auxiliary constraints, can be found easily. In our situation, we take the information about the derivatives of $E[\gamma]$ at $\gamma$ and construct an effective noninteracting system in grand canonical equilibrium whose grand potential functional $\Omega_0[\gamma]$ has the same functional derivative in $\gamma$. The minimum of this energy surface is found by a diagonalization of the effective Hamiltonian and an occupation of the new \abref{ON}s according to the Fermi-Dirac distribution. The resulting \abref{eq}-\abref{1RDM} will then serve as the starting point for the subsequent iteration. This method automatically incorporates the constraints on the \abref{ON}s and \abref{NO}s and we will not have to apply subsequent orthonormalizations and the like. The success of this scheme, of course, relies on the similarity of the energy surfaces of $E[\gamma]$ and $\Omega_0[\gamma]$. 

We will now proceed to derive the variational equations, guiding the determination of $\gamma$.

\subsection{Effective Hamiltonian}\label{sec.num.heff}
The effective noninteracting system is constructed such that the derivatives of the interacting as well as of the noninteracting functional (Eqs. \eq{eq.e} and \eq{eq.gp-ni}) coincide.
\begin{align}
  \frac{\Omega_0[\gamma]}{\delta\gamma(x,x')}&=\frac{\delta E[\gamma]}{\delta\gamma(x,x')}
\end{align}
Because of the possibility of pinned states, this equation does not have to be fulfilled exactly. Therefore, as mentioned before, our minimization routine may not reach the exact minimum but will approach it arbitrarily closely.
The effective Hamiltonian in spatial representation then becomes
\begin{multline}
  h^{eff}[\gamma](x,x')=t[\gamma](x,x')+v_{ext}(x,x')+\\
  \mu\delta(x-x')+1/\beta\sigma[\gamma](x,x')+v_{w}[\gamma](x,x')\label{eq.num.veff}.
\end{multline}
The functional derivatives are given by
\begin{align}
  v_{w}[\gamma](x,x')&=\frac{\delta W[\gamma]}{\delta\gamma(x,x')}\\
  \sigma[\gamma](x,x')&=\frac{\delta S_0[\gamma]}{\delta \gamma(x,x')},
\end{align}
We want to use the chain rule for the functional derivative. We therefore need the derivatives of the \abref{ON}s and \abref{NO}s with respect to $\gamma$. They can be obtained using first-order perturbation theory, yielding
\begin{align}
  \frac{\delta n_k}{\delta \gamma(x',x)}&=\phi_k^*(x')\phi_k(x)\\
  \frac{\delta \phi_k(y)}{\delta \gamma(x',x)}&=\sum_{l\neq k}\frac{\phi_l^*(x')\phi_k(x)}{n_k-n_l}\phi_l(y)\label{eq.num.var.2}\\
  \frac{\delta \phi^*_k(y)}{\delta \gamma(x',x)}&=\sum_{l\neq k}\frac{\phi_k^*(x')\phi_l(x)}{n_k-n_l}\phi^*_l(y)\label{eq.num.var.3}.
\end{align}
In the following, it will be useful to work in the basis of \abref{NO}s. An arbitrary function $g(x,x')$ is then represented by $g_{ij}$, where
\begin{align}
  g_{ij}&=\int dxdx'\phi_i^*(x)g(x,x')\phi_j(x').
\end{align} 
The matrix elements $h^{eff}_{ij}$ of the effective Hamiltonian then become
\begin{multline}
  h^{eff}_{ij}=\delta_{ij}\left(\frac{\partial E[\gamma]}{\partial n_i}+\mu+\frac{\sigma_i}{\beta}\right)+\\
  \frac{1-\delta_{ij}}{n_i-n_j}\int dy\left(\frac{\delta E[\gamma]}{\delta\phi_i(y)}\phi_j(y)-\frac{\delta E[\gamma]}{\delta\phi^*_j(y)}\phi^*_i(y)\right)\label{eq.num.heff.ij},
\end{multline}
where the entropic contribution $\sigma_i$ is given by
\begin{align}\label{eq.num.sigma_i}
  \sigma_{i}&=\frac{\partial S_0[\gamma]}{\partial n_i}=\ln{\left(\frac{1-n_i}{n_i}\right)}.
\end{align}
The offdiagonal elements are exactly the ones Pernal \cite{Pernal.2005} derived in her approach for the derivation of an effective potential for \abref{RDMFT}. They are also simply related to the ones Piris and Ugalde \cite{Piris_Ugalde.2009} introduced in their method for an orbital minimization. It has to be noted, however, that in our approach the diagonal elements are not free to choose but are determined by the thermodynamic ensemble. The temperature of the Kohn-Sham system has no physical meaning and can be varied to influence the convergence behaviour of the minimization routine. If $\beta$ was small, i.e. if the corresponding effective temperature was high, the diagonal part of $\hat H^{eff}$ will be bigger compared to the offdiagonal parts. Therefore, after a diagonalization of $\hat H^{eff}$, the orbitals will change less. When considering the change in \abref{ON}s, one can investigate the limit of $\beta\rightarrow0$. The diagonal of $\hat H^{eff}$ will then just contain the entropic contribution $\sigma_i$. A solution of Eq. \eq{eq.num.heff.ij} will then leave the \abref{ON}s invariant. We will further investigate the behaviour of our self-consistent minimization scheme for small $\beta$ later on in this work.

In the following, we will show how the concept of a temperature tensor greatly enhances the adaptability of the Kohn-Sham system which will improve the performance of the minimization procedure.

\subsection{Temperature tensor}\label{sec.num.tensor}
\begin{figure}[t!]
\centering
\subfigure[First \abref{ON}, $\alpha_1=50$]
{
  \includegraphics[width=.465\columnwidth]{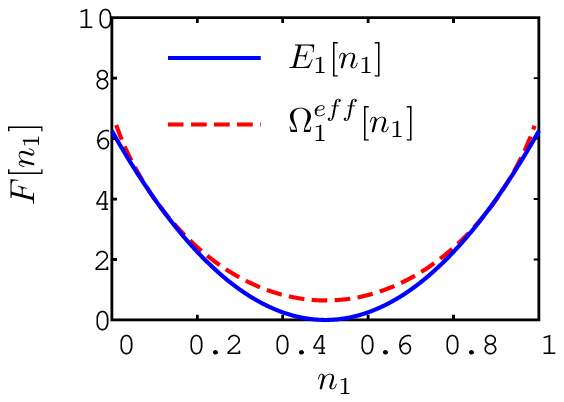}
}\hfill
\subfigure[Second \abref{ON}, $\alpha_2=1$]
{
  \includegraphics[width=.465\columnwidth]{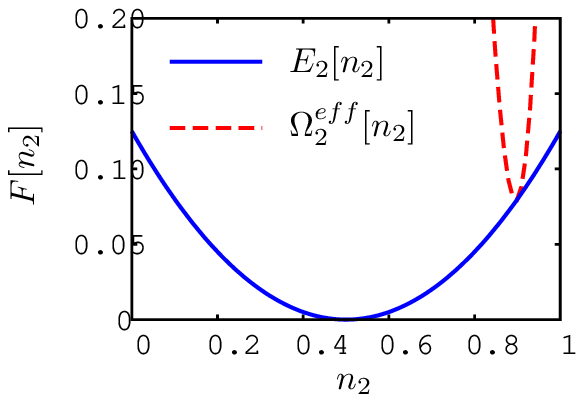}
}
  \caption{Projected energy surfaces for the model of Eq. \eq{eq.num.model} with $\alpha_1=50,\alpha_2=1,\beta=0.11$.}
  \label{fig.num.model.1}
\end{figure}

To understand how the concept of a temperature tensor improves the performance of the minimization scheme, the following considerations will be helpful. In a self-consistent minimization scheme, for a given \abref{1RDM}, we construct a known (noninteracting) functional whose first derivative coincides with the one from the interacting functional. For a fixed $\beta$, the parameter $\mu$ is determined by the requirement of particle number conservation. $\beta$ can then be varied to modify how narrow the noninteracting energy surface should be. However, second derivatives with respect to the \abref{ON}s may differ quite substantially and a value of $\beta$ which describes the energy surface w.r.t. one \abref{ON} well might describe others quite badly. A simple example is the following quadratic two-state model functional $E[n_1,n_2]$ without orbital dependence.
\begin{align}
  E[n_1,n_2]&=\frac{\alpha_1}{2}(n_1-0.5)^2+\frac{\alpha_2}{2}(n_2-0.5)^2\label{eq.num.model}\\
  &=E_1[n_1]+E_2[n_2].
\end{align}
The choice of $\alpha_1=50$ and $\alpha_2=1$ leads to $h^{eff}_{11}=\veps_1=-0.225+\mu$ and $h^{eff}_{22}=\veps_2=0.00450+\mu$ in Eq. \eq{eq.num.heff.ij}.
The corresponding projected grand potential surfaces from Eq. \eq{eq.num.gp-ni-exp} are plotted in Figure \ref{fig.num.model.1} for $\beta=0.11$.
As one can see, the choice of $\beta=0.11$ models the first energy surface quite well, but the second one fails to be reproduced. One would like to have some sort of state-dependent $\beta_i$ which can be related to the second derivatives. However, before one can use such a construct, one has to confirm that it corresponds to an energy-surface whose minimum can easily be found.

We are now going to show that this is possible by a slight variation of the definition of grand canonical ensembles. We consider the following generalized \abref{SDO}-grand potential functional
\begin{align}
  \mf{G}[\hat D]=\tr{\hat D(\mc{\hat B}(\hat H-\mu\hat N)+\ln\hat D)},
\end{align}
where $\mc{\hat B}$ is an arbitrary hermitean operator on the Fock-space. The same proof as in \cite{Mermin.1965} now leads to the following variational principle
\begin{align}
  \mf{G}[\hat D]&\geq\mf{G}[\hat D_{eq}],
\end{align}
where the equality is only fulfilled if $\hat D=\hat D_{eq}$, with
\begin{align}
  \hat D_{eq}&=e^{-\mc{\hat B}(\hat H-\mu\hat N)}/Z_{eq}\label{eq.num.sdo.eq.beta}\\
  Z_{eq}&=\tr{e^{-\mc{\hat B}(\hat H-\mu\hat N)}}.
\end{align}
For a noninteracting Hamiltonian and a $\mc{\hat B}$ for which $[\mc{\hat B},\hat H]=0$ the Fermi Dirac relation reads
\begin{align}
  n_i&=\frac{1}{e^{\beta_i(\veps_i-\mu)}+1}\\
  \veps_i-\mu&=\frac{1}{\beta_i}\ln\left(\frac{1-n_i}{n_i}\right)\label{eq.num.ei.ni.beta},
\end{align}
where $\beta_i$ denotes the i-th eigenvalue of $\mc{\hat B}$. This leads to the following expression for the grand potential
\begin{align}
  \Omega_0[\gamma]&=\sum_i\Big(n_i(\veps_i-\mu)+\nn\\
  &\hspace*{15mm}\frac{1}{\beta_i}(n_i\ln n_i+(1-n_i)\ln(1-n_i))\Big)\label{eq.num.gp.f2}\\
  &=\sum_i\Omega_{0i}[n_i,\beta_i]
\end{align}
\begin{figure}[t!]
\subfigure[First \abref{ON}, $\beta=.5$]
{
  \includegraphics[width=.465\columnwidth]{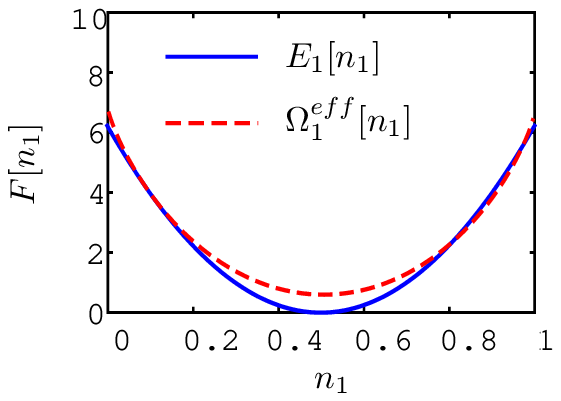}
  }
\hfill
\subfigure[Second \abref{ON}, $\beta_2=5.55$]
{
  \includegraphics[width=.465\columnwidth]{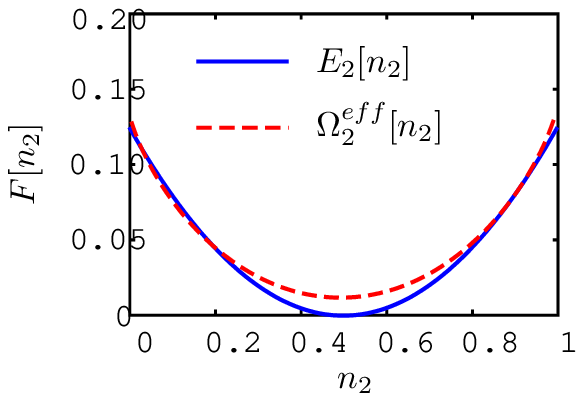}
}
  \caption{Projected energy surfaces for the model of Eq. \eq{eq.num.model} with $\alpha_1=50,\alpha_2=1$. The choice of $\eta=0.5$ leads to $\beta_1=0.11$ and $\beta_2=5.55$.}
  \label{fig.num.model.2}
\end{figure}
Where in the case of a scalar temperature we just had one parameter to construct our effective noninteracting system, we now have one for each \abref{ON}. A straightformard utilization of this freedom would be to let the second derivatives of the energy functional with respect to the \abref{ON}s of the interacting functional and the noninteracting one be proportional to each other.
\begin{align}
  \beta_i&=\eta\frac{\partial ^2 S_0[\gamma]}{\partial n_i^2}\Big/\frac{\partial ^2 E[\gamma]}{\partial n_i^2}\label{eq.num.eta}\\
  &=\eta\frac{1}{n_i(1-n_i)}\left(\frac{\partial ^2 E[\gamma]}{\partial n_i^2}\right)^{-1}\label{eq.num.beta.i},
\end{align}
where $\eta$, the proportionality factor, is the only global parameter. In our model (Eq. \eq{eq.num.model}), this yields
\begin{align}
  \beta_i&=\frac{\eta}{\alpha_i}\frac{1}{n_i(1-n_i)}.
\end{align}
$\eta=1$ lets the second derivatives of interacting and noninteracting functional be equal whereas an increase (decrease) of $\eta$ leads to a spreading (compression) of the noninteracting energy surface. As can be seen from Figure \ref{fig.num.model.2}, with a good choice of $\eta$ (in our model $\eta=0.5$) one can reproduce the different energy surfaces simultaneously.

\begin{figure}[t!]
  \includegraphics[width=\columnwidth]{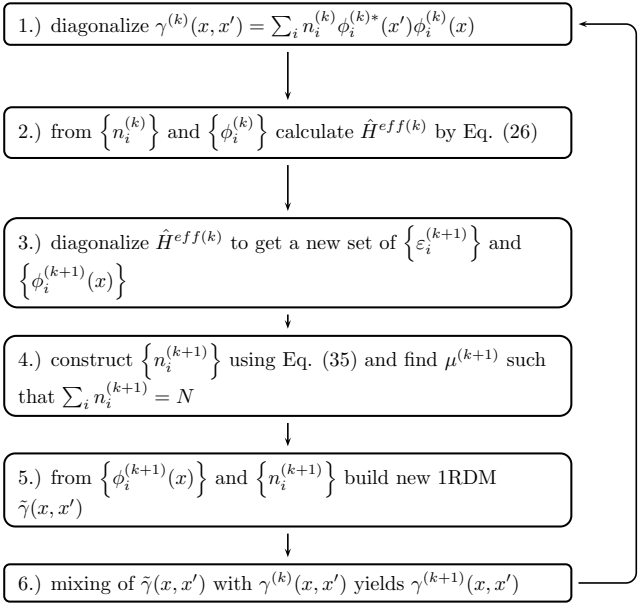}
  \caption{Self-consistent minimization scheme in \abref{FT-RDMFT}}\label{fig.num.sc}
\end{figure}

We can now construct a self-consistent scheme for the minimization of $E[\gamma]$ which we sketch in Figure \ref{fig.num.sc}. A mixing of \abref{1RDM}s is straightforward, because $\Gamma^N$ is a convex set.

\subsection{Small step investigation}\label{sec.num.small-steps}
We showed in the previous considerations that one can employ the Kohn-Sham system in \abref{FT-RDMFT} to construct a self-consistent minimization scheme. However, this does not ensure that an application of this scheme will actually lead to a minimum of the functional. This is a common problem of minimization schemes, but in the following we are going to show that for small steplengths our method will definitely lead to a decrease of the value of the functional under consideration. As we argued before, choosing a smaller $\beta$ will lead to smaller changes in \abref{ON}s and \abref{NO}s. Starting from a given \abref{1RDM} $\gamma$, we therefore apply first-order perturbation theory to get the modified \abref{1RDM} $\gamma'$. By the virtue of Eq. \eq{eq.num.heff.ij}, $\gamma$ leads to the effective Hamiltonian $\hat H^{eff}$. A diagonalization under the assumptions of first-order perturbation theory then yields the following new eigenvalues $\veps_i'$ and eigenstates $\phi_i'$.
\begin{align}
  \veps_i'&=h^{eff}_{ii}\\
  \phi_i'(x)&=\phi_i(x)+\sum_{j\neq i}\frac{h^{eff}_{ji}}{\veps_i-\veps_j}\phi_j(x)
\end{align}
The new \abref{ON}s, resulting from our modified eigenenergies, become
\begin{align}
  n_i'&=\frac{1}{1+e^{\beta_i(\veps_i'-\mu-\Delta_\mu)}},
\end{align}
where one had to introduce the chemical potential correction $\Delta_\mu$ to ensure particle number conservation. With Eqs. \eq{eq.num.heff.ij} and \eq{eq.num.sigma_i} one gets
\begin{align}\label{eq.num.niprime}
  n_i'&=\frac{n_i}{n_i+(1-n_i)e^{\beta_i\left(\frac{\partial E}{\partial n_i}-\Delta_\mu\right)}}.
\end{align}
Expanding Eq. \eq{eq.num.niprime} in orders of $\beta_i$ and retaining the leading contribution, we get
\begin{align}
  \delta n_i&=n_i'-n_i\\
  &=\beta_in_i(n_i-1)\left(\frac{\partial E[\gamma]}{\partial n_i}-\Delta_\mu^{(0)}\right)\label{eq.num.dni}.
\end{align}
This result is very similar to the steepest-descent method with an additional factor of $n_i(n_i-1)$. This additional term tries to keep the \abref{ON}s in the allowed set $0<n_i<1$. $\Delta_\mu^{(0)}$ can now be found by the requirement of particle number conservation,
\begin{align}
  \Delta_\mu^{(0)}&=\frac{\sum_i\beta_in_i(n_i-1)\frac{\partial E[\gamma]}{\partial n_i}}{\sum_i\beta_in_i(n_i-1)}\label{eq.num.rdmks.dmu}.
\end{align}
The overall change in the \abref{1RDM} up to first order in $\beta_i$ is then given by
\begin{align}
  \Delta\gamma_{ij}&=\gamma'_{ij}-\delta_{ij}n_i\\
  &=\delta_{ij}\delta n_i+(1-\delta_{ij})\frac{n_i-n_j}{\veps_i-\veps_j}h^{eff}_{ij}.
\end{align}
The energy changes accordingly as
\begin{align}
  \Delta E&=\int dxdx'\frac{\delta E[\gamma]}{\delta\gamma(x,x')}\Delta\gamma(x',x)\\
  &=\sum_{ij}\frac{\delta E[\gamma]}{\delta\gamma_{ij}}\Delta\gamma_{ji}\\
  &=\underbrace{\sum_i \delta n_i\frac{\partial E[\gamma]}{\partial n_i}}_{\Delta E_1}+\underbrace{\sum_{i\neq j}\frac{n_i-n_j}{\veps_i-\veps_j}|h^{eff}_{ij}|^2}_{\Delta E_2}\label{eq.num.ksrdm.gp}.
\end{align}
We see that the energy change $\Delta E$ seperates into two parts. One is determined by the change in \abref{ON}s, and one comes from the change in \abref{NO}s. In the following we are going to investigate these two different contributions seperately.

\subsubsection{Occupation number contribution}
We will now show that the first term in Eq. \eq{eq.num.ksrdm.gp}, which is due to the change in \abref{ON}s, is negative for appropriately small steplengths.
\begin{align}
  \Delta E_1&=\sum_i \delta n_i\frac{\partial E[\gamma]}{\partial n_i}\\
  &=\sum_i \beta_in_i(n_i-1)\left(\frac{\partial E[\gamma]}{\partial n_i}-\Delta_\mu^{(0)}\right)\frac{\partial E[\gamma]}{\partial n_i}
\end{align}
For brevity, we introduce $c_i=\frac{\beta_in_i(n_i-1)}{\sum_i\beta_in_i(n_i-1)}$. Using Eq. \eq{eq.num.rdmks.dmu} then leads to
\begin{align}
  \Delta E_1&=\left(\sum_j\beta_jn_j(n_j-1)\right)\nn\\
  &\hspace*{5mm}\sum_ic_i\left(\left(\frac{\partial E[\gamma]}{\partial n_i}\right)^2-\left(\sum_kc_k\frac{\partial E[\gamma]}{\partial n_k}\right)^2\right)\\
  &=\left(\sum_j\beta_jn_j(n_j-1)\right)\nn\\
  &\hspace*{18mm}\sum_ic_i\left(\frac{\partial E[\gamma]}{\partial n_i}-\sum_kc_k\frac{\partial E[\gamma]}{\partial n_k}\right)^2.
\end{align}
Because every \abref{ON} $n_i$ fulfills $0<n_i<1$ and every $\beta_i$ is greater 0, this leads to the conclusion
\begin{align}
  \Delta E_1&\leq0.
\end{align}

\subsubsection{Natural orbital contribution}
We can now turn to the second term in Eq. \eq{eq.num.ksrdm.gp} which represents the energy change due to the change in \abref{NO}s.
\begin{align}
  \Delta E_2&=\sum_{i\neq j}\frac{n_i-n_j}{\veps_i-\veps_j}|h^{eff}_{ij}|^2
\end{align}
By using Eq. \eq{eq.num.ei.ni.beta} this transforms to
\begin{align}
  \Delta E_2&=\sum_{i\neq j}\frac{n_i-n_j}{\frac{1}{\beta_i}\ln\left(\frac{1-n_i}{n_i}\right)-\frac{1}{\beta_j}\ln\left(\frac{1-n_j}{n_j}\right)}|h^{eff}_{ij}|^2.
\end{align}
We see that for an arbitrary choice of $\beta_i$, we cannot ensure the negativity of $\Delta\Omega_2$. But if we use a constant $\beta$, we get
\begin{align}
  \Delta E_2&=\beta\sum_{i\neq j}\frac{n_i-n_j}{\ln\left(\frac{n_j(1-n_i)}{n_i(1-n_j)}\right)}|h^{eff}_{ij}|^2,
\end{align}
which is nonpositive for $0<n_i<1$:
\begin{align}
  \Delta E_2&\leq0.
\end{align}

We have shown that for small enough $\beta_i$, the \abref{ON} change will always decrease the grand potential, regardless of wheather one chooses a constant temperature or a temperature tensor. When considering changes in the \abref{NO}s, one has to fall back to constant temperature to ensure a decrease of the functional value. We will use these findings in our numerical implementation of the minimization scheme as demonstrated later on in this work.

\subsection{Convergence measures}\label{sec.num.conv}
We have now all the neccessary tools at hand to iteratively minimize a functional $E[\gamma]$. We need, however, some measures to judge if a calculation is converged. There are two main reasons why using the energy itself as convergence measure is disadvantageous. Firstly, often the calculation of derivatives is not accurate and consequently a derivative-based minimization may lead to a fixpoint where $E[\gamma]$ is not minimal. This leads to a sign change of the convergence measure and implementing the strict decrease of energy as a requirement of the minimization procedure will then lead to a starting point dependent result. Secondly, because the true minimal energy is not known, one would have to judge convergence from the change in $E[\gamma]$ after iterating the minimization routine, i.e. a small change in $E[\gamma]$ indicates a relative closeness to the real minimum. This might pose a problem if the minimum of the energy-surface, as defined by the derivatives of $E[\gamma]$, is very shallow or, worse, if the minimization procedure leads to a slow approach to the minimum. An example for such a situation is discussed in the summary.

Because of these problems, we would rather use a strictly positive convergence measure which goes to 0 if the 1RDM approaches the minimum of the energy-surface, as defined by the derivatives of $E[\gamma]$. We will establish our choice of convergence measures on the following two observations.
\begin{itemize}
  \item In the minimum, the derivatives with respect to the \abref{ON}s will be equal for unpinned states.
  \item In the minimum, the effective Hamiltonian $\hat H^{eff}$ will be diagonal.
\end{itemize}
The first observation allows us to define a convergence measure $\chi^2_n$ for a minimization with respect to the \abref{ON}s:
\begin{align}
  \chi^2_n&=\frac{1}{N_{unpinned}}\sum_i^{N_{unpinned}}\left(\frac{\partial E}{\partial n_i}-\overline\mu\right)^2\label{eq.num.chi2n}\\
  \overline\mu&=\frac{1}{N_{unpinned}}\sum_i^{N_{unpinned}}\frac{\partial E}{\partial n_i}.
\end{align}
The second statement leads to the following definition of $\chi^2_\phi$ as a convergence measure for a minimization with respect to the \abref{NO}s:
\begin{align}
  \chi^2_\phi&=\frac{1}{N-1}\frac{\sum_{i\neq j}^{N}|h^{eff}_{ij}|^2}{\sum_i\veps_i^2}.
\end{align}
If a minimization is converging, both measures should approach 0.

\section{Example}\label{sec.num.samples}
We test the self-consistent procedure for the case of solid LiH at zero temperature by using the FP-LAPW code Elk \cite{Elk}. The exchange-correlation energy will be modelled by the  $\alpha$ functional $E_{xc}^\alpha[\gamma]$, as introduced in \cite{Sharma_al.2008}. The interaction functional $W[\gamma]$ is then given as a sum of the classical Hartree energy functional $E_H[\gamma]$ and the exchange-correlation functional $E_{xc}^{\alpha}[\gamma]$.
\begin{multline}
  E_{xc}^{\alpha}[\gamma]=-\frac12\sum_{ij}n_i^\alpha n_j^\alpha\int dxdx' \\
  w(x,x') \phi_i^*(x')\phi_i(x)\phi_j^*(x)\phi_j(x')
\end{multline}
We choose this functional because it exhibits several properties making it difficult to be minimized. It will lead to several fully occupied, i.e. pinned states. Therefore, as argued before, there is no noninteracting system at finite temperature reproducing this 1RDM exactly but there will be one leading to an equilibrium 1RDM arbitrarily close. The minimization of the $\alpha$ functional is therefore a good test for the minimization scheme leading to boundary minima on $\Gamma^N$. Furthermore, the $\alpha$ functional exhibits divergencies in the derivatives w.r.t. the \abref{ON}s for $n_i\rightarrow0$. If, in the minimum, there are \abref{ON}s close to 0 (and there will be if one considers enough \abref{NO}s) this might lead to convergence problems of the minimization. 

We will now investigate the performance of the self-consistent minimization scheme w.r.t. \abref{ON}- and \abref{NO}-convergence. It turned out in the course of our investigations that a mixing of 1RDMS, according to point 6 in Figure \ref{fig.num.sc}, does not improve our results and we therefore abstain from it.

\subsubsection{Occupation number minimization}
\begin{figure*}%
\centering
  \subfigure[Energy: 1 k-point]
  {
    \label{fig.num.rdmks.on.k1.e}
    \includegraphics[width=0.48\textwidth]{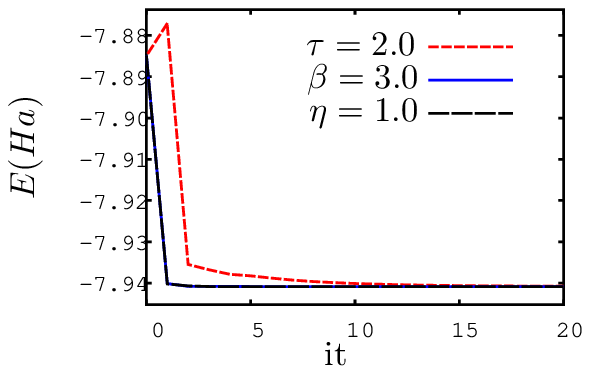}
  }\hfill
  \subfigure[\abref{ON}-Variance: 1 k-point]
  {
    \label{fig.num.rdmks.on.k1.chi2}
    \includegraphics[width=0.48\textwidth]{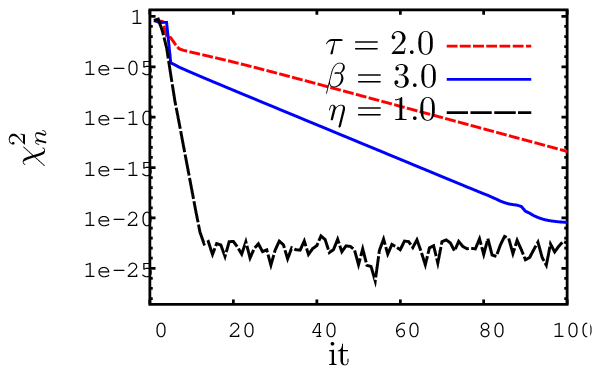}
  }\\
  \subfigure[Energy: 2x2x2 k-points]
  {
    \label{fig.num.rdmks.on.k2.e}
    \includegraphics[width=0.48\textwidth]{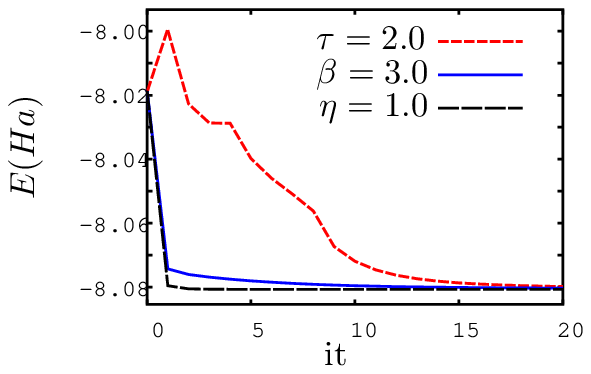}
  }\hfill
  \subfigure[\abref{ON}-Variance: 2x2x2 k-points]
  {
    \label{fig.num.rdmks.on.k2.chi2}
    \includegraphics[width=0.48\textwidth]{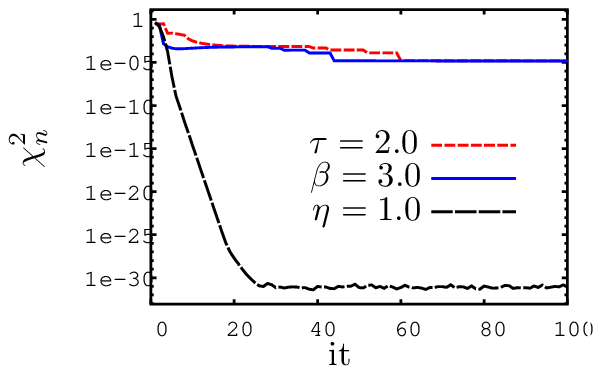}
  }\\
  \subfigure[Energy: 3x3x3 k-points]
  {
    \label{fig.num.rdmks.on.k3.e}
    \includegraphics[width=0.48\textwidth]{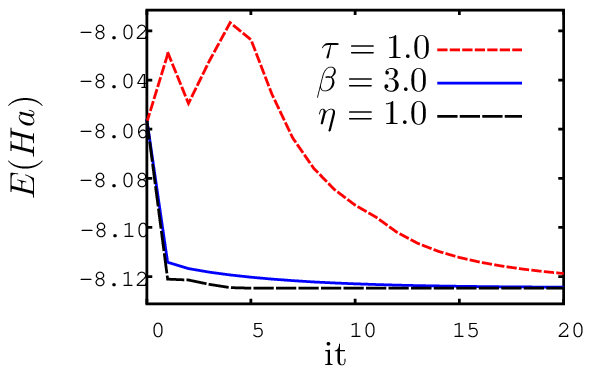}
  }\hfill
  \subfigure[\abref{ON}-Variance: 3x3x3 k-points]
  {
    \label{fig.num.rdmks.on.k3.chi2}
    \includegraphics[width=0.48\textwidth]{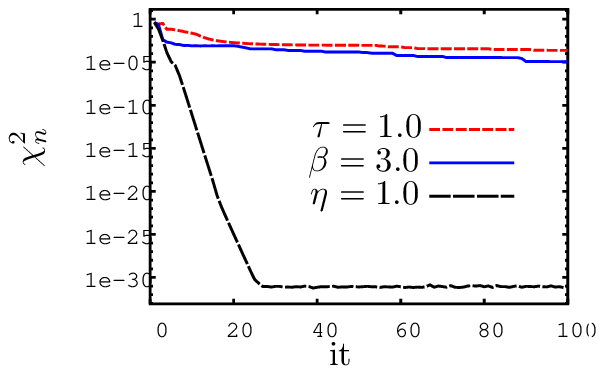}
  }
  \caption[\abref{ON}-minimization of the $\alpha$ functional applied to LiH]{Energy $E$ and \abref{ON}-convergence measure $\chi_n$ for \abref{ON} minimizations of the $\alpha$ functional, with $\alpha=0.565$, applied to LiH. The red, short dashed lines stand for a steepest-descent minimization, the blue, solid ones for a \abref{sc}-Kohn-Sham minimization with constant $\beta$, and the black, long dashed ones for a \abref{sc}-Kohn-Sham minimization with adaptive $\beta_i$. $\tau$ denotes the parameter value for $taurdmn$ in Elk, whereas $\beta$ and $\eta$ are defined via Eqs. \eq{eq.gp-ni} and \eq{eq.num.eta}.}
  \label{fig.num.rdmks.on}%
\end{figure*}

We have minimized the $\alpha$ functional for $\alpha=0.565$ with three methods. First, we have used the steepest-descent method, as implemented in Elk. The second method is the self-consistent \abref{FT-RDMFT} minimization with constant $\beta$, and finally we have employed a temperature tensor $\beta_i$ of the form of Eq. \eq{eq.num.beta.i} with parameter $\eta$. In all three methods, we chose all parameters to achieve fastest convergence. The results, which are shown in Figures \ref{fig.num.rdmks.on.k1.e}-\ref{fig.num.rdmks.on.k3.chi2}, show that both self-consistent Kohn-Sham minimizations lead to a faster convergence than steepest-descent. A dramatic improvement in the speed of convergence is achieved by employing a temperature tensor. The slow decrease of $\chi^2_n$ in Figures \ref{fig.num.rdmks.on.k2.chi2} and \ref{fig.num.rdmks.on.k3.chi2} for the steepest-descent and constant-$\beta$ methods can be attributed to the following fact. For these two methods, the \abref{ON}s, which will be pinned at the equilibrium, approach their final values quite slowly. Therefore, their derivatives contribute to $\chi^2_n$ via Eq. \eq{eq.num.chi2n} even after several iterations.

% ...

\subsubsection{Full minimization}
\begin{figure}[t!]
  \includegraphics[width=0.8\columnwidth]{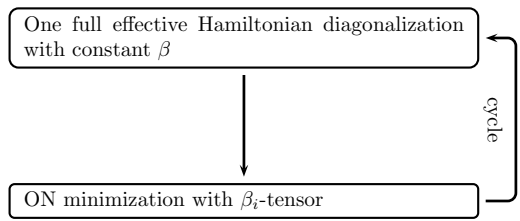}
  \caption{Full minimization scheme}\label{fig.num.full}
\end{figure}

\begin{figure*}%
\centering
  \subfigure[Energy: 1 k-point]
  {
    \label{fig.num.rdmks.full.k1.e}
    \includegraphics[width=0.48\textwidth]{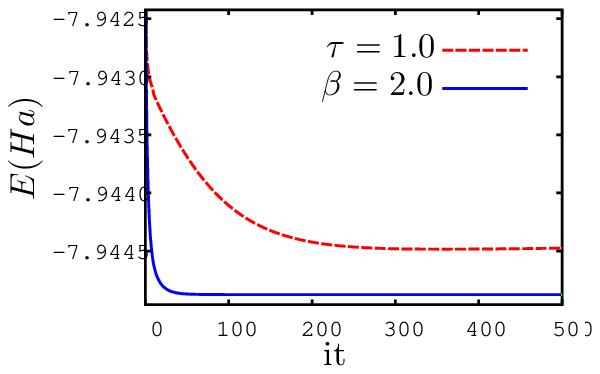}
  }\hfill
  \subfigure[\abref{NO}-Variance: 1 k-point]
  {
    \label{fig.num.rdmks.full.k1.chi2}
    \includegraphics[width=0.48\textwidth]{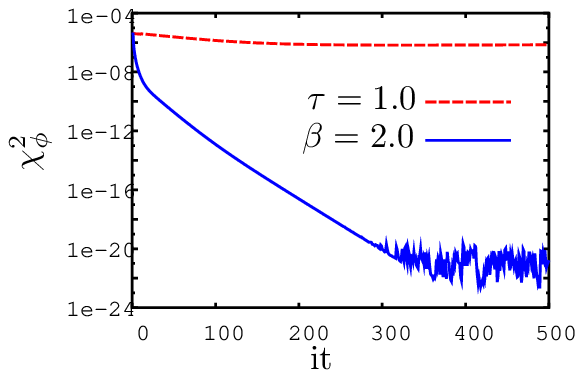}
  }\\
  \subfigure[Energy: 2x2x2 k-points]
  {
    \label{fig.num.rdmks.full.k2.e}
    \includegraphics[width=0.48\textwidth]{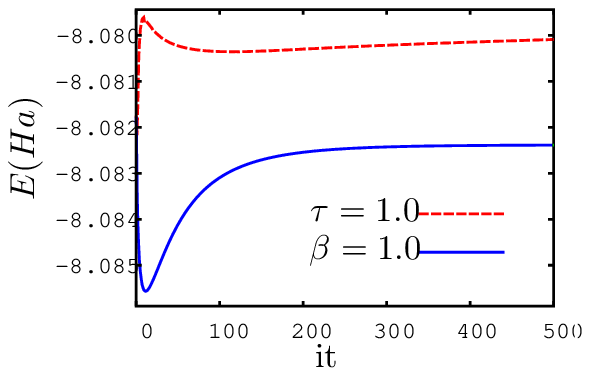}
  }\hfill
  \subfigure[\abref{NO}-Variance: 2x2x2 k-points]
  {
    \label{fig.num.rdmks.full.k2.chi2}
    \includegraphics[width=0.48\textwidth]{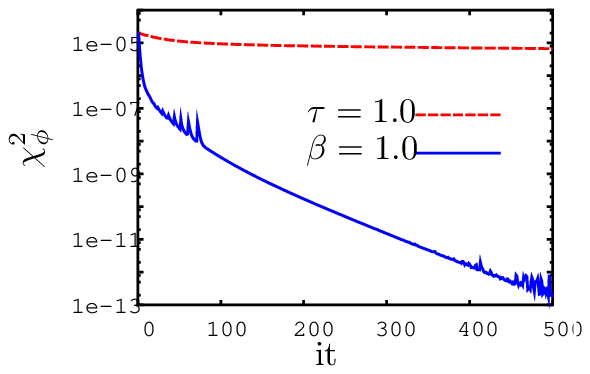}
  }\\
  \subfigure[Energy: 3x3x3 k-points]
  {
    \label{fig.num.rdmks.full.k3.e}
    \includegraphics[width=0.48\textwidth]{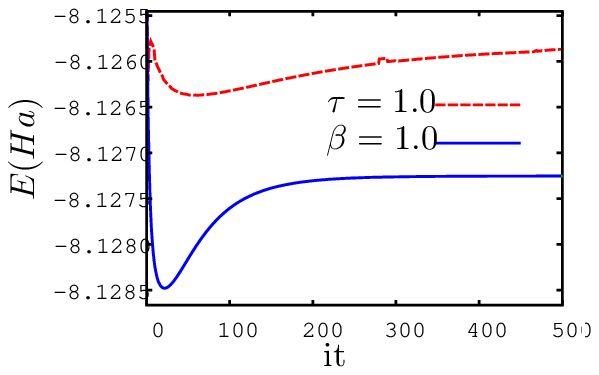}
  }\hfill
  \subfigure[\abref{NO}-Variance: 3x3x3 k-points]
  {
    \label{fig.num.rdmks.full.k3.chi2}
    \includegraphics[width=0.48\textwidth]{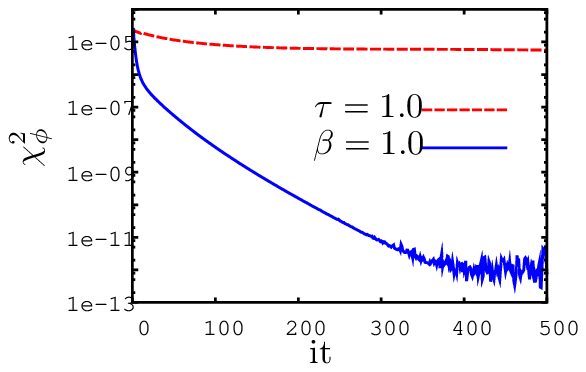}
  }

    \caption[\abref{NO}-minimization of the $\alpha$ functional applied to LiH]{Energy $E$ and \abref{NO}-convergence measure $\chi_\phi$ for \abref{NO} minimizations of the $\alpha$ functional, with $\alpha=0.565$ applied to LiH. Both variables are plotted against the number of \abref{NO} changes. After each change in \abref{NO} there follows a complete \abref{ON} minimization. The red, dashed lines stand for a steepest-descent minimization whereas the blue, solid ones depict a \abref{sc}-Kohn-Sham minimization with constant $\beta$. $\tau$ denotes the parameter value for $taurdmc$ in Elk, whereas $\beta$ is defined via Eq. \eq{eq.gp-ni}. The increase of energy is due to the fact that the derivatives are calculated only approximately.}
  \label{fig.num.rdmks.full}%
\end{figure*}

We can now turn to the problem of minimizing $E[\gamma]$ with respect to both \abref{ON}s and \abref{NO}s. We find that the overall performance of this full minimization is greatly improved by introducing a \abref{ON}-minimization after every \abref{NO}-minimization step (see Figure \ref{fig.num.sc}). Because we have seen in the previous section that this can be done very efficiently, this increases the runtime of a full minimization run only negligibly. The deeper reason for the improvement of the convergence by inclusion of an \abref{ON}-minimization is the following: It typically happens that two states $\phi_i$ and $\phi_j$ have similar eigenvalues in $\hat H^{eff}$ but considerably different \abref{ON}s. A diagonalization of $\hat H^{eff}$ then yields a strong mixing between these states. If the \abref{ON}s were not updated, one might be led away from the minimum of the grand potential functional. A subsequent \abref{ON}-minimization remedies this problem and assigns the optimal \abref{ON} for each \abref{NO}. We show a sketch of the full minimization scheme in Figure \ref{fig.num.full}. An application of this scheme to LiH then leads to the results depicted in Figure \ref{fig.num.rdmks.full}. Again, we see a tremendous increase in speed and accuracy for the self-consistent Kohn-Sham minimization scheme compared to the steepest-descent method. The steepest-descent method shows a very slow convergence, which can be attributed to the orthonormalization of \abref{NO}s. The increase of the energy curves in Figures \ref{fig.num.rdmks.full.k2.e} and \ref{fig.num.rdmks.full.k3.e} is due to the approximative nature of the derivatives.



\cleardoublepage

\cleardoublepage

\section{Summary and outlook}

\begin{figure}[t]
  \includegraphics[width=\columnwidth]{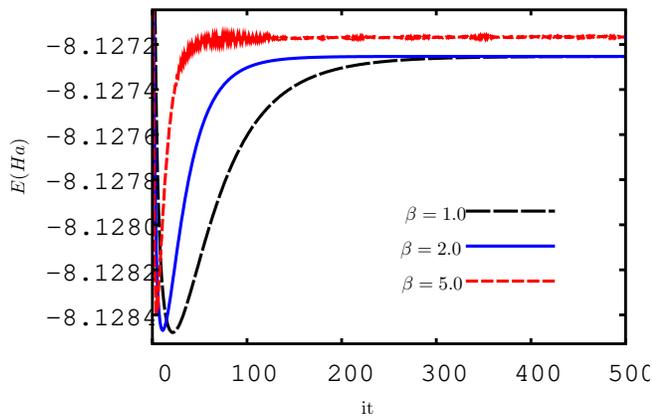}
  \caption{Energy convergence for different effective temperatures.}\label{fig.num.rdmks.three.e}
\end{figure}

\begin{figure}[t]
  \includegraphics[width=\columnwidth]{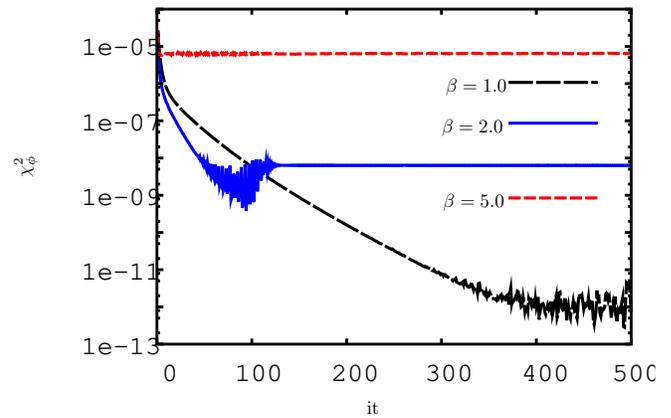}
  \caption{\abref{NO}-convergence for different effective temperatures.}\label{fig.num.rdmks.three.chi}
\end{figure}

In the present work, we have introduced a self-consistent Kohn-Sham minimization scheme in the theoretical framework of \abref{RDMFT}. We have defined measures which allow us to judge the convergence of a calculation without having to resort to the energy. We could show that this self-consistent procedure is superior in many respects compared to the steepest-descent method, especially considering a minimization w.r.t. the \abref{NO}s. The important parameter in the minimization scheme is the effective temperature $\beta$ and the speed of convergence crucially depends on it. In Figures \ref{fig.num.rdmks.three.e} and \ref{fig.num.rdmks.three.chi} we show the behaviour of the minimization scheme for three different choices of $\beta$. $\beta=1$ represents the optimal value, i.e. the value for which the convergence measure $\chi_\phi^2$ decreases the fastest. We see that the energy reaches its fixpoint after approximately 300 iterations. An increase of $\beta$ to $\beta=2$ seemingly speeds up the energy convergence, but from $\chi_\phi^2$ one can see that after about 100 iterations the minimization fails to diagonalize $h^{eff}$ any further. The changes in the \abref{1RDM}, whose amplitudes are determined by $\beta$, become too big and the \abref{1RDM} jumps around the fixpoint of the energy. Without considering $\chi_\phi^2$, this would have been difficult to detect which illustrates the importance of a convergence measure which is independent of the energy value. One might argue that this choice of $\beta$ still leads to a fixpoint very close to the optimal one, but this cannot be ensured for all problems and all choices of $\beta$ and therefore has to be seen in the actual example as rather accidental, i.e. fortunate. A further increase of $\beta$ to $\beta=5$ then exposes this problem more dramatically. The energy apparently reaches a fixpoint. But this fixpoint is considerably above the optimal one. Just having the energy at hand, this would have been difficult to detect. But $\chi_\phi^2$ directly shows that the minimization is far from being converged.

One important feature, which can be extracted from Figures \ref{fig.num.rdmks.three.e} and \ref{fig.num.rdmks.three.chi}, is that all three parameters lead to a similar energy vs. iteration curve. Apparently, a minimization-run with $\beta$ being too big is able to lead to the vicinity of the fixpoint. An utilization of this fact would now be to use an adaptive $\beta$ rather than a constant one. One could start with a big $\beta$ till the energy does not change anymore and than decrease $\beta$ until $\chi_\phi^2$ surpasses the convergence threshold.

We expect that our successful demonstration of an efficient minimization scheme in RDMFT will support the investigation and development of functionals and therefore encourage further work in this field of research.

\bibliographystyle{apsrev}

\end{document}